\newcommand{\ba}{\begin{array}}
\newcommand{\ea}{\end{array}}
\newcommand{\bean}{\begin{eqnarray*}}
\newcommand{\eean}{\end{eqnarray*}}

\newcommand{\bmx}[1]{\left(\begin{array}{*{#1}{c}}}
\newcommand{\emx}{\end{array}\right)}
\newcommand{\bmxw}[1]{\renewcommand{\arraystretch}{2}\left(\begin{array}{*{#1}{c}}}
\newcommand{\bmxww}[1]{\renewcommand{\arraystretch}{2.5}\left(\begin{array}{*{#1}{c}}}
\newcommand{\bdet}[1]{\renewcommand{\arraystretch}{1.2}
	\left|\begin{array}{*{#1}{c}}}
\newcommand{\edet}{\end{array}\right|\renewcommand{\arraystretch}{1}}
\newcommand{\beq}{\begin{equation}}
\newcommand{\eeq}{\end{equation}}
\newcommand{\bea}{\begin{eqnarray}}
\newcommand{\eea}{\end{eqnarray}}

\newcommand{\ditem}[1]{\item[$\diamond$]}

\newcommand{\bit}{\begin{itemize}}
\newcommand{\eit}{\end{itemize}}

\newcommand{\eab}{\begin{eqnarray}}
\newcommand{\eae}{\end{eqnarray}}

\documentclass[aps,prb,twocolumn,superscriptaddress]{revtex4-1}
\usepackage{subfig}
\usepackage{graphicx}
\usepackage{amssymb}
\usepackage{color} 
\usepackage{amsmath}

\begin{document}


\title{Duality at Lattice Scale around the Takhtajan-Babujain Point in the Bilinear Biquadratic Spin-1 Chain}


\author{Kuang-Ting Chen}
\affiliation{Lawrence Berkeley National Laboratory, Berkeley, CA 94720}


\date{\today}

\begin{abstract}
In the literature, it is known that near the Takhtajan-Babujain point of the bilinear-biquadratic spin-$1$ chain, the low energy theory is captured by three Majorana fermions with a mass sign change, or three Ising chains near the transition. In this paper, we study how the spin-$1$, the Ising spins, and the fermions are related at lattice scale. We construct the spin-$1$ operator out from the Ising spins and fermions, showing that when the Ising chains are in the ordered phase, the spin-$1$ chain is in the dimer phase, and when the Ising chains are in the paramagnetic phase, the spin-$1$ chain is in the Haldane phase. Through this mapping we can see how unit cell doubling in the dimer phase and the boundary spin-$\frac12$ zero mode in the Haldane phase take place, even though in the Ising transition neither phenomena exist. 

\end{abstract}


\maketitle


\section{Introduction}
There has been a long history of the research on the bilinear-biquadratic spin-1 chain\cite{tsvelik,chubukov1991,shelton1996,kitazawa1999,ivanov2003,liu2012}, characterized by the Hamiltonian 
\beq\label{bbspinh}
H=J \sum_i\big( S_i\cdot S_{i+1}-b(S_i\cdot S_{i+1})^2\big).
\eeq 
It contains quite a few points of interest: at $b=0$ it is the Heisenberg chain, which is first conjectured to be gapped by Haldane\cite{haldane1983}. At $b=-\frac13$, the system is gapped and the ground state wave function is known exactly\cite{aklt}. At $b=1$, the Takhtajan-Babujian (TB) point, the system is exactly solvable by the Bethe ansatz\cite{Babujian1983}. The low energy states at this point are gapless with linear dispersion, and the effective theory thus possesses conformal symmetry. With the conformal symmetry, all the relevant operators are known, and the effective low energy theory near the TB point is three Majorana fermions with either a positive or negative mass term (and a marginal current-current interaction), as shown by Tsvelik\cite{tsvelik}.

Recently, a renewed interest arose, due to the realization that the gapped Haldane phase is a symmetry protected topological phase (SPT)\cite{xie2013}. The TB point is thus a transition between a traditional symmetry breaking phase (the dimer phase) and a SPT. However, from the point of view of the effective low energy theory of three Majorana chains, the two phases are only distinguished by the sign of the fermion mass term. There are no associated features that distinguish between the two phases, such as zero modes at the two ends of the chain in the Haldane phase, or unit cell doubling in the dimer phase.

To observe such features, it is necessary to relate the lattice spin to the low energy fermions. In the literature, such relations are given between the primary operators in the conformal theory and the coarse-grained version of the lattice spins, i.e., local spin averages and staggered spin averages.\cite{shelton1996,kitazawa1999} While these relations are useful to derive correlation functions from the low energy theory, they contain no information about the degeneracies at the boundary or unit cell doubling, since neither does exist in continuum. 

On the other hand, in Ref. \cite{liu2012}, the authors use slave fermions with Guzwiller projections to numerically study the bilinear-biquadratic chain around the TB point. They also have qualitatively argued how the resulting $Z_2$ gauge theory coupled with fermions can explain unit cell doubling and have the correct ground state degeneracy. The drawback of this approach is that the fermions are always strongly coupled via the constraint, which is expressed as the coupling to the $Z_2$ gauge field. While one can argue about when the $Z_2$ gauge field would be deconfined, even then it is not straightforward to claim that the mean field theory of the fermions is the low energy effective theory.

In this paper, we take a different route comparing to the two approaches above. We start from three Majorana chains and assume they go through the Ising transition at the same time. We then construct a spin-1 operator using the fermion and Ising spin operators, showing the resulting spin-1 chain are in the dimer/Haldane phase when the three chains are in the ordered/disordered phase. From this construction we can then bridge our understanding from the low energy effective theory toward a lattice scale duality. Unlike the transverse Ising model, however, when we express the bilinear-biquadratic Hamiltonian using the Majorana fermions and Ising spins, it is still not exactly solvable.

\section{A Review of 1d Ising Chain}
For our purpose, it is very important to understand the lattice 1d Ising chain first. Even though it is well known we still provide a brief but self-contained introduction here.

The 1d transverse Ising model is defined by the following Hamiltonian:
\beq
H=-\sum_i \sigma^z_i\sigma^z_{i+1}-g\sum_i \sigma^x_i;
\eeq
$\sigma^z$ and $\sigma^x$ are Pauli matrices and satisfy $(\sigma^z_i)^2=(\sigma^x_j)^2=1$; $\{\sigma^x_i,\sigma^z_j\}=2\sigma^x_i\sigma^z_j(1-
\delta_{ij})$. When $|g|<1$, the chain is in the ordered phase with spontaneous symmetry breaking $\langle\sigma^x\rangle\neq 0$. When $|g|>1$, the chain is in the paramagnetic phase when the spins align themselves with the ``magnetic field" in the $x$ direction. $|g|=1$ is the critical point when the system is gapless.

There is a duality between $g=a$ and $g=1/a$: we define the dual spin 
\beq\label{muz}
\mu^z_i=\prod_{j\leq i} \sigma^x_j;
\eeq
\beq\label{mux}
\mu^x_i=\sigma^z_{i}\sigma^z_{i+1}.
\eeq
It is straightforward to check that $(\mu^x_i)^2=(\mu^z_i)^2=1$, $[\mu^z_i,\mu^x_j]=0$ for $i\neq j$ and $\{\mu^z_i,\mu^x_i\}=0$. In terms of $\mu$-spin the Hamiltonian then becomes
\beq
H=-\sum_i \mu^x_{i}-g\sum_i\mu^z_{i-1}\mu^z_{i}.
\eeq
The duality relation transforms the two terms in the Hamiltonian into each other; the ordered phase of the $\sigma$-spin is therefore the paramagnetic phase of the $\mu$-spin and vice versa. It is not as important for our purpose here, but we mention in passing that the ground state degeneracy difference is of global nature, since the duality relation Eq. \ref{mux} restricts the Hilbert space of $\mu$-spin to have $\prod\mu^x=1$, if we take periodic boundary conditions. 

\begin{figure}
\centering
    \subfloat[We use red dots to represent the $\sigma$-spins.]{
        \includegraphics[width=4cm]{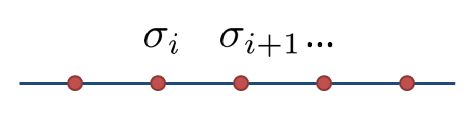}
    }
    \subfloat[The $\mu$-spins, represented by green squares, are naturally placed between the $\sigma$-spins due to Eq. \ref{muz} and \ref{mux}.]{
        \includegraphics[width=4cm]{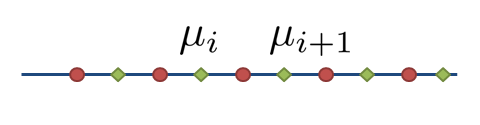}        
    }
    
    \subfloat[The Majorana fermions $\chi$ and $\eta$ are formed by multiplying the nearby $\sigma$-spin and $\mu$-spin together as in Eq. \ref{chi} and \ref{eta}.]{
        \includegraphics[width=4cm]{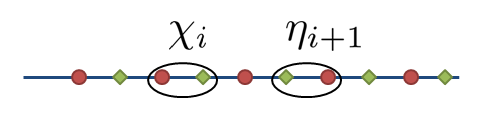}
    }
    \subfloat[The degrees of freedom on the chain can be completely described by the two Majorana fermions, represented by black and white squares.]{
        \includegraphics[width=4cm]{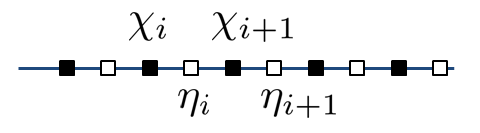}        
    }
    
    \subfloat[The two phases of the Ising chain can be thought of as two different ways to pair the Majorana fermions.]{
        \includegraphics[width=4cm]{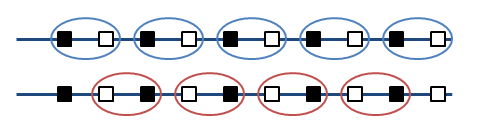}        
    }
    \caption{\label{isingpic}Some simple diagrams illustrating the relation between different local operators of the Ising chain.}
\end{figure}

This Hamiltonian is also exactly solvable. To see this, one needs to translate it into a fermionic problem. If we define
\beq\label{chi}
\chi_i\equiv i\sigma^z_i\mu^z_i,
\eeq
and
\beq\label{eta}
\eta_i\equiv \mu^z_{i}\sigma^z_{i+1},
\eeq
from the fact that $\{\mu^z_i,\sigma^z_j\}=0$ for $j\leq i$ and all other $\sigma^z$'s and $\mu^z$'s commute, we can see that $\{\chi_i,\eta_j\}=0$ for any $i,j$ and $\{\chi_i,\chi_j\}=\{\eta_i,\eta_j\}=0$ for $i\neq j$, as they all contain exactly one pair of $\sigma$ and $\mu$ from each operator that anticommutes. In addition, $\chi_i^2=\eta_i^2=1.$ $\chi_i$ and $\eta_i$ are therefore Majorana fermions. The relations of the local operators defined on the chain are illustrated in Fig. \ref{isingpic}.

In terms of the fermions, the Hamiltonian now reads
\beq
H=\sum_ii\chi_i\eta_i+g\sum_ii\eta_i\chi_{i+1}.
\eeq
Evidently, the Hamiltonian is now quadratic in fermionic operators and can be diagonalized using the usual fermionic creation and annihilation operators formed by pairing the Majorana fermions, in momentum space basis. In terms of the resulting fermions, the total number of fermions is only conserved modulo two, and the two phases are both superconductors. When we look at the exemplary Hamiltonians of the two phases from the point of view of Majorana fermions, it is then obvious that the two phases are distinguished by how the Majorana modes are paired together. In the ``conventional" superconductor phase the Majorana modes pair with each other on site. In the ``topological" superconductor phase the Majorana modes pair between neighboring sites, creating one unpaired Majorana mode at each end. Right at the critical point, the system becomes translation invariant with respect to the Majorana fermions, and the resulting gapless excitation is a single branch of Majorana fermion, with the central charge $c=1/2$, half the degrees of freedom of the complex fermion.

The important takeaway points for us from the model are:
\begin{enumerate}
\item The two phases in the 1d transverse Ising model can be understood as the condensation of either the $\sigma$-spin or the $\mu$-spin, with  $\langle\sigma_z\rangle\neq 0$ or $\langle\mu_z\rangle\neq 0$.
\item The two phases can also be understood as different pairing patterns between the Majorana fermions. The two pairing patterns are related by half a lattice shift of the Ising spin. 
\item In terms of the fermions, this half lattice shift of the duality transformation signals the fact that the transition is between a normal phase and a topological phase.
\end{enumerate}     
\section{The Spin-1 Chain}
In this section we shall construct a spin-1 operator per site out from three Ising/Majorana chains. Let us label the chains by $x$, $y$,$z$, and call the fermions on the $i$ chain $\chi^i$ and $\eta^i$. We start from the obvious construction:
\beq
S^a=\frac{s^{\chi a}+s^{\eta a}}2\equiv\frac{i}{4}\epsilon_{abc}(\chi^b\chi^c+\eta^b\eta^c);
\eeq 
$a$, $b$, $c$ runs through $x$, $y$, and $z$. We suppress the site index since all operators are on the same site. $S^a$ is a sum of two independent spin-$\frac12$ $s^\chi$ and $s^\eta$. The Majorana fermion bilinears $s^{\chi a}\equiv\frac{i}{2}\epsilon_{abc}(\chi^b\chi^c)$are spin-$\frac12$ operators because they have the same algebra as Pauli matrices:
\beq
(s^{\chi a})^2=1; \;[s^{\chi a},s^{\chi b}]=2\epsilon_{abc}s^{\chi c};\;\{s^{\chi a},s^{\chi b}\}=2\delta_{ij}.
\eeq
$s^\eta$ is similarly defined and has the same properties. The two sets of spin evidently commute with each other.

We can easily understand the construction when we think about the states with three fermions per site. In total we have eight states; when the total fermion number is 0 or 3, they are singlets, and when the fermion number is 1 or 2 they form triplets. Since $s^\chi$ and $s^\eta$ commutes with the fermion parity operator the four states from fermion number 0 and 2 as well as 1 and 3 each contain a spin-1 representation and a spin-0 representation, as we would expect from two independent spin-$\frac12$s. If we restrict ourselves to the sector with fixed fermion number 1 or 2, we get a spin-1 operator per site.

However, this construction does not yield the correct behavior of the spin-1s when we assume the underlying Majorana chains are transitioning between the two Ising phases. First of all, in one of the phases where the Majorana fermions bound within site, the total fermion number per site is either 0 or 3. This state thus does not survive the projection as proposed above. In the other phase the ground state can survive the projection, but the projection most certainly will alter the dynamics of the system. Furthermore, if we consider spin correlations, there will be no anti-ferromagnetic correlations, as the fermion system is invariant when translated by the lattice constant.

This is related to the fact that there are different primary operators in the $SU(2)_2$ Wess-Zumino-Witten theory capturing long range spin and staggered spin correlations.\cite{witten1984,affleckHaldane,kitazawa1999,shelton1996} The primary operators that correspond to long range spin operators are fermion bilinears $\epsilon_{abc}\xi^a\xi^b$ of scaling dimension $(\frac12,\frac12)$ where $\xi's$ are the three Majorana fermions in the continuum theory. The staggered spin corresponds to the $(\frac3{16},\frac3{16})$ primary operators that can be expressed as products of the Ising operators. They are not locally expressible in terms of the fermions\cite{tsvelik,shelton1996}.

We therefore are led to consider a similar construction on the lattice to add in the antiferromagnetic correlations. We will also need to use the Klein factors explicitly. Klein factors are constant global factors that anti-commute with each other. They are attached to every chain to make sure that the fermions formed on different chains also anti-commute with each other. With an odd number of chains, it is possible to take $K_n\propto\prod_{i=1}^{n-1} K_i$, which anti-commutes with all other $K_i$. In terms of the Ising spins and the Klein factors, the spin operators read
\begin{align}
s^{\chi a}_i&=\frac{i}{2}\sum_{bc}\epsilon_{abc}K_{b}K_{c}\sigma^b_i\mu^b_i\sigma^c_i\mu^c_i;\\
s^{\eta a}_i&=\frac{i}{2}\sum_{bc}\epsilon_{abc}K_{b}K_{c}\mu^b_i\sigma^b_{i+1}\mu^c_i\sigma^c_{i+1}.
\end{align}
Here $\sigma$ and $\mu$ stand for $\sigma^z$ and $\mu^z$ respectively; the $z$ index is suppressed since we are not going to refer to either $\sigma^x$ or $\mu^x$. We have also restored the site index $i$, since not every operator is on the same site. 

Now consider the operator
\beq
\tilde s^{\chi x}_i\equiv K_x\sigma^x_i\mu^y_i\mu^z_i;
\eeq
other components of $\tilde s$ are similarly defined. By explicit calculation, we find that
\begin{align}
[\tilde s^{\chi a}_i,\tilde s^{\chi b}_j]&=2\delta_{ij}\epsilon_{abc}s^{\chi c}_i,\\
[ s^{\chi a}_i,\tilde s^{\chi b}_j]&=2\delta_{ij}\epsilon_{abc}\tilde s^{\chi c}_i,\\
[s^{\eta a}_i,\tilde s^{\chi b}_j]&=0.
\end{align}  
This enables us to define
\beq
s^{\chi a}_{i,\pm}\equiv\frac{s^{\chi a}_i\pm\tilde s^{\chi a}_i}2,
\eeq
which satisfies the SU(2) commutation relationship $[s^{\chi a}_{i,\pm},s^{\chi b}_{j,\pm}]=2\delta_{ij}\epsilon_{abc}s^{\chi c}_{i,\pm}$. In addition, the $+$ and the $-$ spins commute: $[s^{\chi a}_{i,+},s^{\chi b}_{j,-}]=0$. Interestingly, if we ask what is the spin of this operator by calculating $(s^\chi_\pm)^2$, we get
\beq\label{totalschi}
(s^\chi_\pm)^2=\frac14\big(6\pm 6(iK_xK_yK_z\sigma^x\sigma^y\sigma^z)\big);
\eeq 
$x$, $y$, $z$ are chain labels, not labeling Pauli matrices. If we choose the convention such that $K_xK_yK_z=-i$, the $+$ and $-$ spins are either of spin $\frac12$ or 0, depending on the sign of the order parameter when the Ising chains are in the ordered phase, and a mixture of those when the chains are in the disordered phase.

We also define $\tilde s^{\eta}_i$ similarly:
\beq
\tilde s^{\eta x}_i\equiv K_x\sigma^x_{i+1}\mu^y_i\mu^z_i;
\eeq
it is nevertheless important to notice that it is not the dual of $\tilde s^{\chi}$. The similar commutations with $s^\eta$ also follow. We then define $s^{\eta}_\pm$ by
\beq
s^{\eta a}_{i,\pm}\equiv\frac{s^{\eta a}_i\mp\tilde s^{\eta a}_i}2;
\eeq
the opposite sign is there to ensure that
\beq\label{totalseta}
(s^\eta_{i,\pm})^2=\frac14\big(6\pm 6(iK_xK_yK_z\sigma^x_{i+1}\sigma^y_{i+1}\sigma^z_{i+1})\big).
\eeq
In a sense, in this construction we ``fractionalized'' the spin-$\frac12$s $s^\chi$ and $s^\eta$ into two parts, differed by $\tilde s$. They are going to be put on nearby lattice sites so that $\tilde s$ gives rise to staggered spin correlations.

Looking at Eq. \ref{totalschi} and Eq. \ref{totalseta}, we also notice that the total spin of $s^\eta_{i,\pm}$ is determined by the same factors as $s^\chi_{i+1,\pm}.$ This then implies that the sum $(s^\eta_{i,\pm}+s^\chi_{i+1,\mp})$ is always a spin-$\frac12$. It is therefore desirable to put two of such sums on nearby sites, so that on every site there are now two spin-$\frac12$s and they sum up to give us our spin-$1$ operator.  

Last but not the least, we notice that in order for $\tilde s$ to consistently be the staggered spin, not only should we put $s_+$ and $s_-$ on nearby sites, but if we put $s_{i,+}$ on site $i$, we also need to put $s_{i+1,-}$ on site $(i+1)$. This assignment will ensure that a slowly varying $\tilde s$ contributes to the staggered spin.

\begin{figure}
\centering
    \subfloat[The red dots and the green squares are the $\sigma$- and $\mu$-spins as before. We have circled the combination that give rise to $s^\chi$, $s^\eta$, $\tilde s^\chi$, and $\tilde s^\eta$. The other components of the spins are formed similarly with the corresponding chains.]{
        \includegraphics[width=6cm]{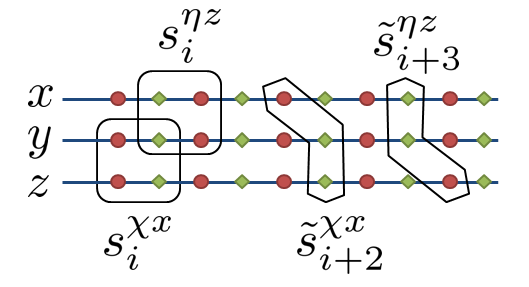}
    }
    
    \subfloat[$s^{\chi/\eta}_\pm$ is the average of $s^{\chi/\eta}$ and $\tilde s^{\chi/\eta}$. Notice how the product of $s^{\chi/\eta}$ and $\tilde s^{\chi/\eta}$ is proportional to the product of $\sigma$'s.]{
        \includegraphics[width=8cm]{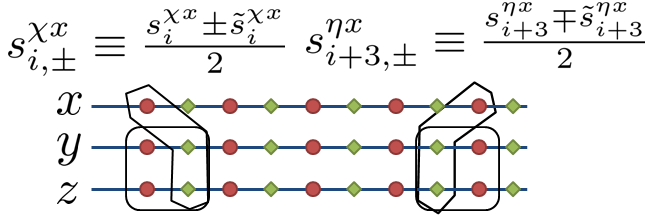}        
    }
    
    \subfloat[the total spin of $s^\chi_{i+1,\pm}$ and $s^\eta_{i,\mp}$ are determined by the same product of $\sigma_{i+1}$'s. Their sum is therefore always a spin-$\frac12$.]{
        \includegraphics[width=6cm]{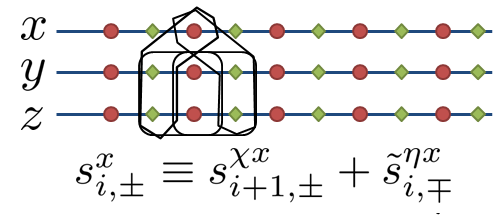}
    }
    
    \subfloat[The spin-$\frac12$s $s_\pm$, represented by the blue and white squares respectively, are arranged to form a spin-$1$ on every site. This particular arrangement ensures that $\tilde s$ carries antiferromagnetic correlations, and will transit from dimer to Haldane phases when the underlying Ising chains go from ordered phase to paramagnetic phase.]{
        \includegraphics[width=8cm]{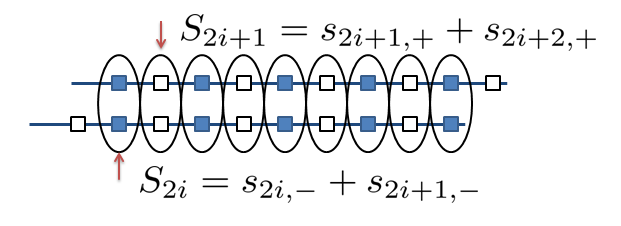}        
    }
    
    \caption{\label{bbchainfig}Some diagrams illustrating the procedure to construct the spin-$1$ operator from the Ising spins of the three chains.}
\end{figure}

In the end, with all the considerations above, we have the on-site spin operator as follows:
\begin{align}\label{main}
S_{2i+1}&=s^\eta_{2i,-}+s^\chi_{2i+1,+}+s^\eta_{2i+1,-}+s^\chi_{2i+2,+};\\
S_{2i}&=s^\eta_{2i-1,+}+s^\chi_{2i,-}+s^\eta_{2i,+}+s^\chi_{2i+1,-}.
\end{align}
This is the central result of the paper. The whole construction is also illustrated pictorially in Fig. \ref{bbchainfig}.

Now we take a look at how the spin-$1$ chain looks like when the underlying Ising/fermion chains are in the two phases:

When the Ising chains are $\sigma$-ordered, the fermions pair within site. Let us suppose that only $+$-spins survive $(iK_xK_yK_z\sigma^x\sigma^y\sigma^z=1)$; in this case, $s^\chi_{i,+}$ will form a singlet with $s^\eta_{i,+}$. We then find that $S_{2i-1}$ form a singlet with $S_{2i}$. This is the dimer phase, with unit cell doubling. If the $\sigma$-spins order in the opposite direction, then only $S_-$s remain, but the fermions still pair on-site. This will lead to $S_{2i}$ forming a singlet with $S_{2i+1}$. This is the other dimer phase, which is the same phase translated by a lattice constant. There is unit cell doubling in both cases, because Eq. \ref{main}, the mapping between the spin-1 operator and the underlying Ising spins and fermions, breaks lattice translation explicitly.

When the Ising chains are $\mu$-ordered, the fermions pair between sites, i.e., $\eta_i$ pair with $\chi_{i+1}$. In this case even though $+$ and $-$ spins both contribute, we see that the latter half of $S_{2i}$ forms a spin-$\frac12$, and pairs with the first half of $S_{2i+1}$ into a singlet. In terms of the spin-$1$s, it fractionalized into two spin-$\frac12$s, and each pairs with the neighboring site on its left or right. This is the Haldane phase. As is well known, at the boundary, a single spin-$\frac12$ is left alone without forming a singlet and that is the origin of the boundary zero mode. In this phase, while Eq. \ref{main} breaks lattice translation symmetry, the singlet forming restores it. All the $S_\pm$s of the same lattice site are contained in the same singlet and it does not matter that the $\pm$ definition is staggered. 

At this point, it is also obvious that we can define a dual spin-$1$ operator by shifting the Ising and fermion operators by half a unit cell in the construction Eq. \ref{main}, i.e., $\sigma_i\rightarrow\mu_i$, $\mu_i\rightarrow\sigma_{i+1}$ and $\chi_i\rightarrow\eta_i$, $\eta_i\rightarrow\chi_{i+1}$. The dual spin will be in dimer phase when the original spin is in Haldane phase, and vice versa. While we have shown how to construct both spin-$1$s using the Ising spins and fermions, it is unclear whether it is possible to express one directly in terms of the other.

\section{Discussion} 
We thus have identified the mapping between the spin-$1$ operator and the Ising spins/fermions, such that when the Ising spins and fermions go through the Ising transition, the spin-$1$ chain will transit between the dimer phase and the Haldane phase. From the construction we can also read out how the spin correlations at low energies can be translated to Ising spin and fermion correlations. 

One issue that we have thus far refrained ourselves from mentioning, is how this construction works at Hamiltonian level. In other words, If we express Eq. \ref{bbspinh} near $b=1$ using Eq. \ref{main}, do we get the transverse Ising Hamiltonian for each chain? 

The answer is no. What we can say is just that the resulting Ising Hamiltonian is in the same universality class as the transverse Ising model. It is not exactly solvable, either. In addition, at the TB point, the duality transform does not match the Hamiltonian into itself. It matches it into itself with some additional irrelevant terms under renormalization group flows.

This is to be expected. Firstly, the Bethe ansatz solution is known not to be equivalent to non-interacting Majorana fermions. This means after our identification of the spin-$1$ operator with a construction from Ising spins and fermions, we should not get the Hamiltonian of the transverse Ising model at transition. Secondly, let us consider how to project the constructed spin to actually be spin-1 (since it is a sum of two spin-$\frac12$s on every site.) At Hamiltonian level it is trivially done by coupling to this spin operator as a whole. The spin-0 component will always decouple. This corresponds to the fact that in the Hamiltonian, the two spin-$\frac12$s that sum up to be the spin-$1$ couple equally to everything else (indeed, the symmetrization of the two spin-$\frac12$s is the projection to spin-1.) However, it is not the case in our construction, when the spin chain is in the Haldane phase. From the Ising Hamiltonian the spin $s^\eta_{i,\pm}$ only couples to $s^\chi_{i+1,\pm}$, but with symmetrization it should also couple to $s^\chi_i$ and $s^\eta_{i-1}$. What we have constructed with the fermions is a state with entangled pairs, but without projection to spin one. The Ising Hamiltonian thus can not be equivalent to any spin-$1$ Hamiltonian in the Haldane phase.

Our construction is therefore not so useful to explore the dynamics of spin-1 chains further away from the transition, as the resulting Hamiltonian we get in terms of Ising spins and fermions are often complicated and strongly interacting. Nevertheless, it is still an excellent example, which illustrates how a proper ``fractionalization" and ``recombination" of degrees of freedom on nearby sites can map an ordinary phase transition into a phase transition between a symmetry breaking phase and a SPT. It would be interesting to consider whether constructions of this kind can be applied in higher dimensions.

\section*{Acknowledgement}
We thank Dunghai Lee for useful discussions. This work was partly supported by Lawrence Berkeley National Laboratory when I was a postdoctoral researcher there.

\bibliography{mybib}

$\;$

\end{document}